\documentclass{aastex}
\usepackage{spr-astr-addons}

\usepackage{url}
\usepackage{graphics}

\RequirePackage{color}

\def\avg#1{\overline{\langle#1\rangle}}

\def\mavg#1{{\langle#1\rangle}_{m}}

\def\vec#1{\mathbf{#1}}
\def\mcol{\multicolumn}
\def\tnm{\tablenotemark}
\def\tnt{\tablenotetext}
\def\ex{$\times$}

\def\msun{M$_{\odot}$}

\begin{document}

\title{Some Properties of the Kinetic Energy Flux and Dissipation in Turbulent Stellar Convection Zones}
\slugcomment{Rome Conference, June 2009}
\shorttitle{Turbulent Stellar Convection}
\shortauthors{Meakin \& Arnett}

\author{Casey A. Meakin} \and \author{W. David Arnett}
\affil{Steward Observatory, University of Arizona, Tucson, AZ, 85721, USA}
\email{casey.meakin@gmail.com}

\begin{abstract}
We investigate simulated turbulent flow within thermally driven stellar convection zones. Different driving sources are studied, including cooling at the top of the convectively unstable region, as occurs in surface convection zones; and heating at the base by nuclear burning. The transport of enthalpy and kinetic energy, and the distribution of turbulent kinetic energy dissipation are studied.  We emphasize the importance of global constraints on shaping the quasi-steady flow characteristics, and present an analysis of turbulent convection which is posed as a boundary value problem that can be easily incorporated into standard stellar evolution codes for deep, efficient convection. Direct comparison is made between the theoretical analysis and the simulated flow and very good agreement is found. Some common assumptions traditionally used to treat quasi-steady turbulent flow in stellar models are briefly discussed. The importance and proper treatment of convective boundaries are indicated.
\end{abstract}



\section{Introduction}

\par  While the equations governing the dynamics of non-magnetized stellar plasma are well known, a fundamental understanding of fully developed turbulent flow remains elusive.  A Reynolds decomposition, whereby the properties of the stellar plasma are separated into mean and fluctuating components $\phi = \avg{\phi} +\phi'$ provides some insight into the problem.  Decomposing the kinetic energy equation (formulated by the product of the velocity and the momentum equation) in this way and taking temporal and angular averages (indicated by the operator $\avg{\cdot}$) results in  \citep{meakin2007b}

\begin{equation}
\label{eq:fke}
\begin{array}{l}
\partial_t\avg{\rho E_K} + \nabla\cdot\avg{\rho E_K \vec{u_0}} =  \\
-\nabla\cdot\avg{\vec{F}_p+\vec{F}_K}  
+\avg{p'\nabla\cdot\vec{u'}}
+\avg{\vec{W}_b}
-\avg{\epsilon_{K}}
\end{array}
\end{equation}

\noindent which is the full non-linear governing equation of interest.  The primary goal of any stellar turbulence theory is to model the terms of this equation, including the rate of buoyancy work $\vec{W}_b = \rho'\vec{u'}\cdot\vec{g}$, the kinetic energy flux $\vec{F}_K = \vec{u'}E_K$, the pressure correlation flux $\vec{F}_p = \vec{u'}p'$, the work done by pressure fluctuations $p'\nabla\cdot\vec{u'}$, and the rate at which kinetic energy $E_K$, is dissipated $\epsilon_K$.  Differential rotation and circulation currents introduce additional sources of turbulence and transport terms. 

\par It is standard practice to ignore or grossly approximate most of these terms in stellar evolution calculations.  For instance, mixing length theory (MLT) ignores $\epsilon_K$, $\vec{F}_p$, $\vec{F}_K$, and $p'\nabla\cdot\vec{u'}$ and approximates the {\em integral} of $\vec{W}_b$ over a {mixing length} as the {\em product} of local properties of the flow.  The time dependence expressed by the left hand side of Eq.~\ref{eq:fke} is also dropped. Though still not widely used, some strides have been made to compensate for these deficiencies through embellished MLT type algorithms, most notably to address the issue of time dependence and the non-local nature of turbulence \citep[e.g.][and references therein]{gough1976,unno1981,eggleton1983,kuhfuss1986,deng2006}.

\par In order to develop a more realistic physical description of stellar turbulence, which is of central importance to modeling stellar pulsation \citep[e.g.][]{belkacem2006,samadi2009}, a better understanding of these non-linear terms is needed.  A powerful method for gaining insight into this physics is analyzing fully non-linear simulation data.
In the following we present a few select models from a new suite of turbulent stellar convection simulations designed to this end and briefly discuss the origin of the kinetic energy flux and its relationship to the kinetic energy dissipation and the large scale topology of the flow.

\begin{table*}
\small
\caption{Selected Simulation Model Parameters\label{tab:models}}
\begin{tabular}{lclcl}
\tableline
\mcol{1}{c}{Model ID} &
\mcol{1}{c}{$\Delta\theta,\Delta\phi$\tnm{\it a}} &
\mcol{1}{c}{Zoning} &
\mcol{1}{c}{t$_{\rm avg}$\tnm{\it b}}&
\mcol{1}{c}{Comments} \\
\mcol{1}{c}{} &
\mcol{1}{c}{[deg.]} &
\mcol{1}{c}{[$n_r\times n_{\theta}\times n_{\phi}$]}&
\mcol{1}{c}{[s]} &
\mcol{1}{c}{}\\
\tableline
{\sf h1}        & 30,30 & 200\ex50\ex50        &  [300, 500] & narrow, static heating profile\\
{\sf h1.z2}   & 30,30 & 400\ex100\ex100    & - & model h1 with moderate resolution increase\\
{\sf h1.z1}   & 30,30 & 800\ex200\ex200    & - & model h1 with high resolution \\
{\sf h3}        & 30,30 & 200\ex50\ex50         & [375, 575] & broad, static heating profile\\
{\sf c1}         & 30,30 & 200\ex50\ex50         & [200, 400] & static top cooling profile \\
\tableline
\end{tabular}
\tnt{a}{The computational domain is centered on the equator so that the 
domain extends $\Delta\theta/2$ degrees above and below the equator.}
\tnt{b}{Provided is the time interval over which averages are performed.}
\end{table*}

\section{Simulation Setup}

\par The initial conditions used in our reactive hydrodynamic simulations 
are based on a 23~\msun~star model which has been evolved with the TYCHO stellar evolution code \citep{arnett2009b} to an age of $\sim2\times 10^6$ yr, at which point oxygen is burning in a shell that overlies a silicon-sulfur-rich core \citep{meakin2007b,meakin2006}.
Variations in the driving source are made in order to study how this impacts the global characteristics of the flow, and in turn how this affects the transport terms.  Three models are presented including two in which heating (by nuclear burning) is present and one in which convection is driven by a cooling region near the top of the convection zone (similar to radiative losses in surface convection zones).  The heating and cooling profiles are presented in Fig~1. The heated and cooled regions ($0.44\lesssim r/10^9 {\rm cm}\lesssim0.64$) are initially nearly adiabatic, and thus neutrally buoyant, while the surrounding layers are stably stratified.
The fully compressible, reactive Euler equations are solved using the PROMPI code \citep{meakin2007b} which is a descendant of the PROMETHEUS piecewise parabolic method (PPM) code \citep{fryxell1989} adapted to parallel computing platforms. We work within the implicit large eddy simulation (ILES) framework to treat the grid-scale dissipation and turbulent stresses \citep[e.g.][]{boris2007,aspden2007,benzi2008}. The sensitivity of our results on resolution are tested within limits of computational cost by a series of higher resolution runs.  A summary of simulation properties is presented in Table~1.

\begin{figure*}
\includegraphics[scale=0.52]{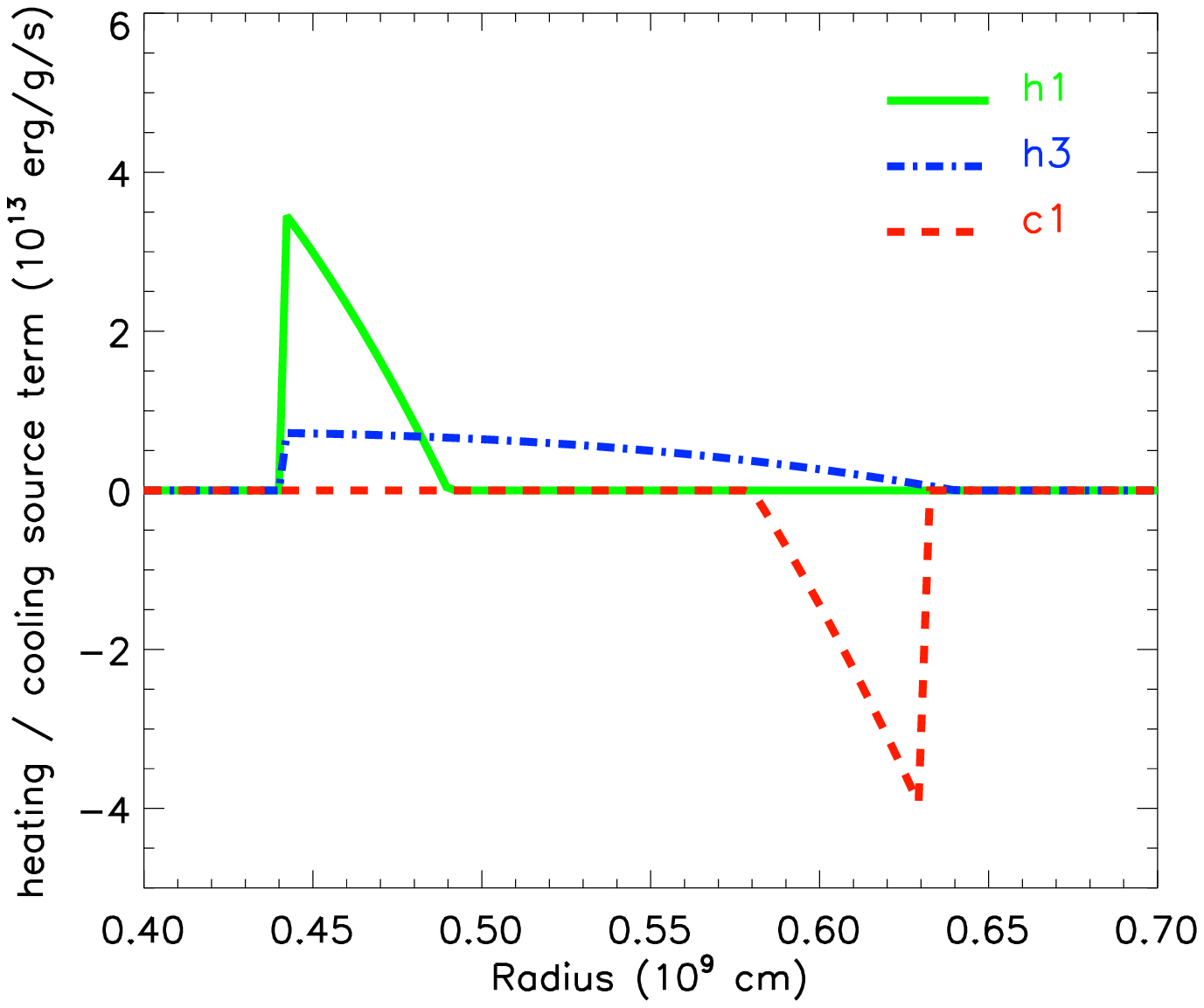}
\includegraphics[scale=0.52]{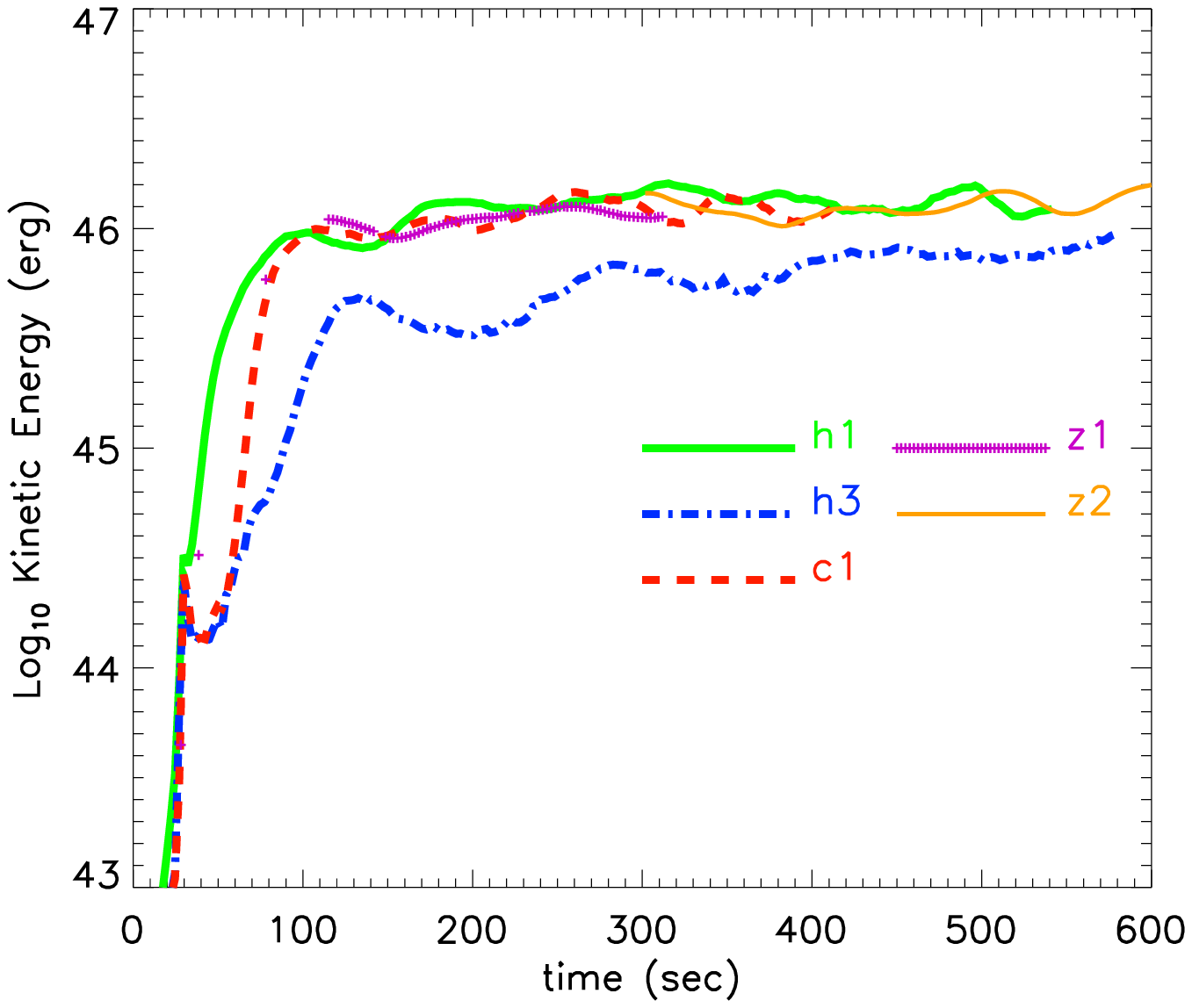}
\caption{(left)  Heating / cooling profiles for the models listed in Table 1. (right) The time evolution of the total kinetic energy in the simulation domain. The two additional high resolution models {\sf h1.z1} and {\sf h1.z2} are indicated by the magenta crosses and the orange line, respectively.}
\end{figure*}

\section{Thermal Relaxation}
\label{sec:therm-balance}

\par The time evolution of the kinetic energy is presented in Fig.~1. The convective turnover time $\tau_c = 2 L/v$ is a little less than $\tau_c \sim$100 s for all of the models studied. After an initial transient comparable to $\tau_c$ the models attain a quasi-steady state. The strong damping present in turbulent flow \citep[see e.g.][]{arnett2009a} ensures that this state is reached within $\sim \tau_c$.  Time averages for analysis are performed over intervals that encompass $\sim 2\tau_c$ and are summarized in Table~1.  

\par The simulated convection is very efficient in all cases and deviates only mildly from an isentropic state. Since the simulated regions are not in thermal balance (there is either a net heating or cooling) the entropy will change over time and the fluxes and flows within the convection zones adjust to establish an isentropic state at each moment.  In this situation the rate of entropy change at any one location is equal to that of the mass averaged rate over the convection zone, $\dot{s} = \mavg{\dot{s}}$, where $\mavg{\cdot}$ indicates a mass weighted average over the turbulent region. From the first law ($\delta Q = dU + \delta W$) and the fundamental thermodynamic relationship ($dU = TdS - \delta W$) 

\begin{equation}
\Big( \frac{\epsilon_n +\epsilon_K}{T} - \frac{dL_{C}/dm}{T}  \Big)  = \mavg{\dot{s}} = \frac{\mavg{\epsilon_n+\epsilon_K}}{\mavg{T}}
\end{equation}

\noindent and the convective luminosity is

\begin{equation}
L_C(m) = \int_{M_0}^{M_0+m}\Big( \epsilon_n +\epsilon_K -  T \mavg{\dot{s}} \Big)dm'
\end{equation}

\noindent where $\epsilon_n$ is the local heating or cooling term (see Fig.~1).

\par The convective flux found from this relationship is compared to the simulation data in Fig~2 for all three models. The good agreement shows  that thermal relaxation is not necessary to study turbulent convection but can be incorporated into the analysis.  A much more important effect than this slow thermal relaxation is the luminosity associated with boundary layer mixing events which is as large as $\sim$40\% of peak in model {\sf c1}.

\par The kinetic energy dissipation $\epsilon_K$ is required to appropriately calculate $\mavg{\dot{s}}$ and is included in standard stellar evolution (i.e., MLT) only implicitly through the structure variable $\nabla=d\ln T/d\ln P$.  The distribution of $\epsilon_K$ throughout a convection zone, however, is intimately related to the resulting kinetic energy flux, which we discuss next.

\begin{figure*}
\includegraphics[scale=0.52]{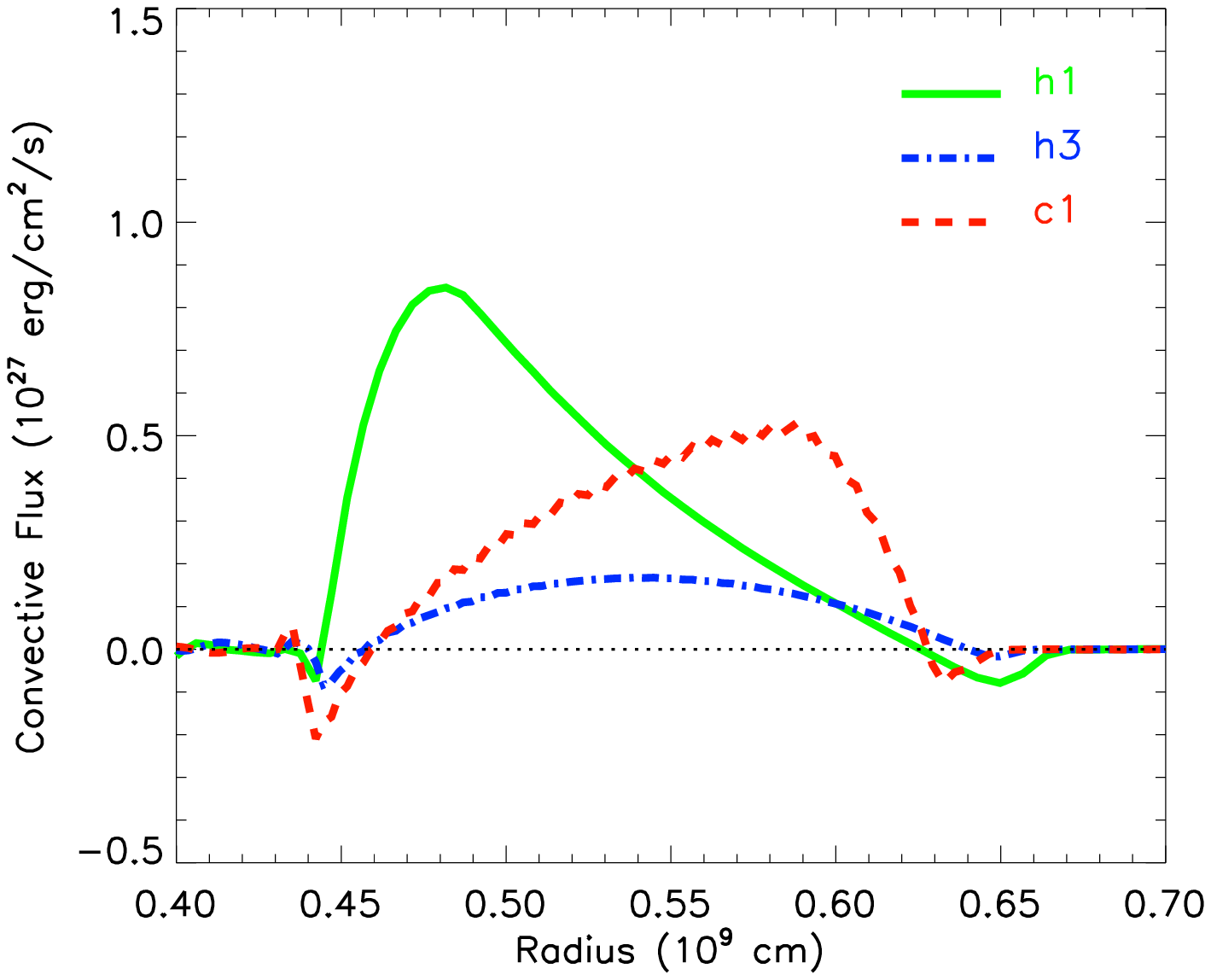}
\includegraphics[scale=0.52]{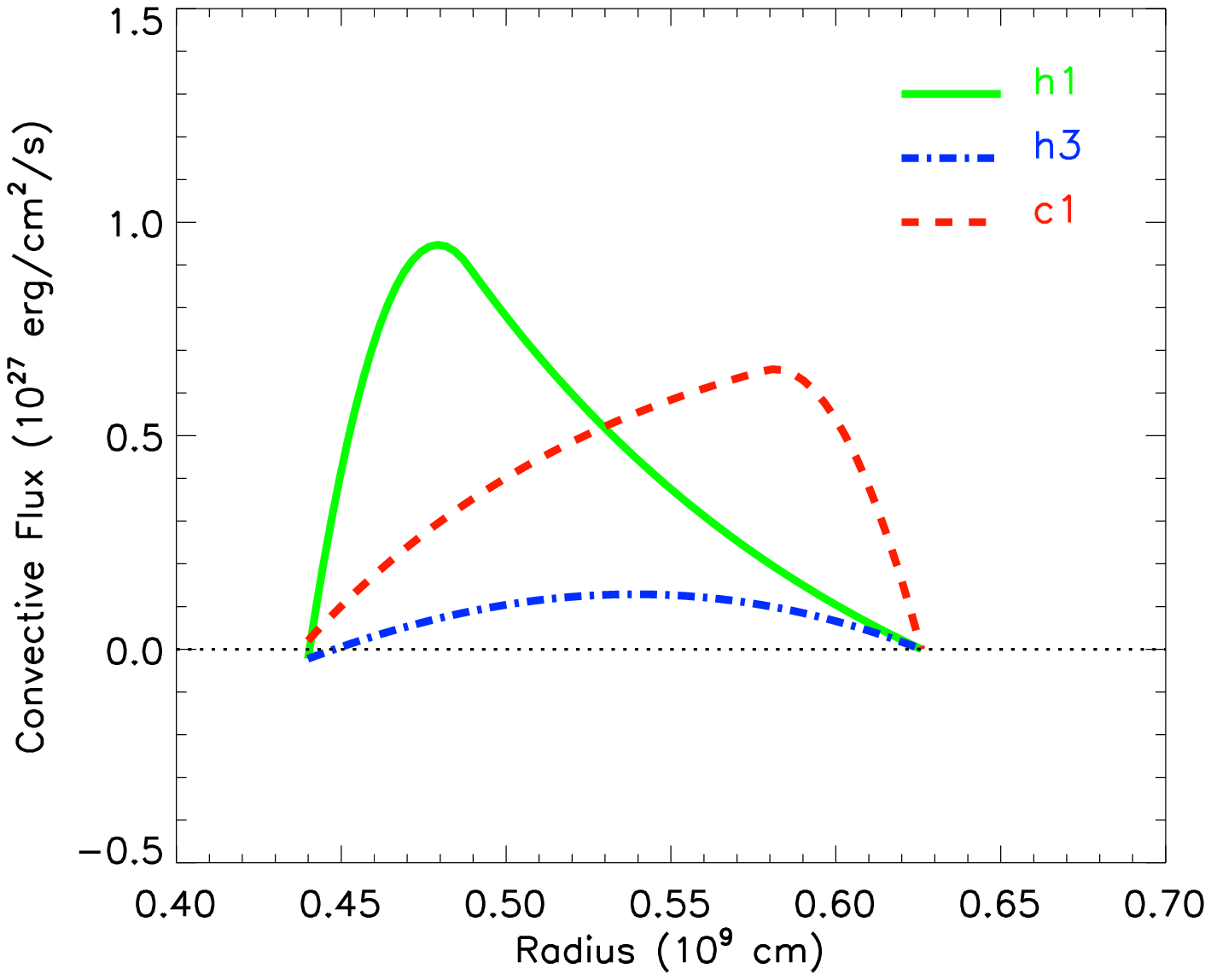}
\caption{Convective flux: (left) time averaged simulation data and 
(right) calculated from the background structure as described by Eq.~3.}
\end{figure*}

\begin{figure*}
\includegraphics[scale=0.52]{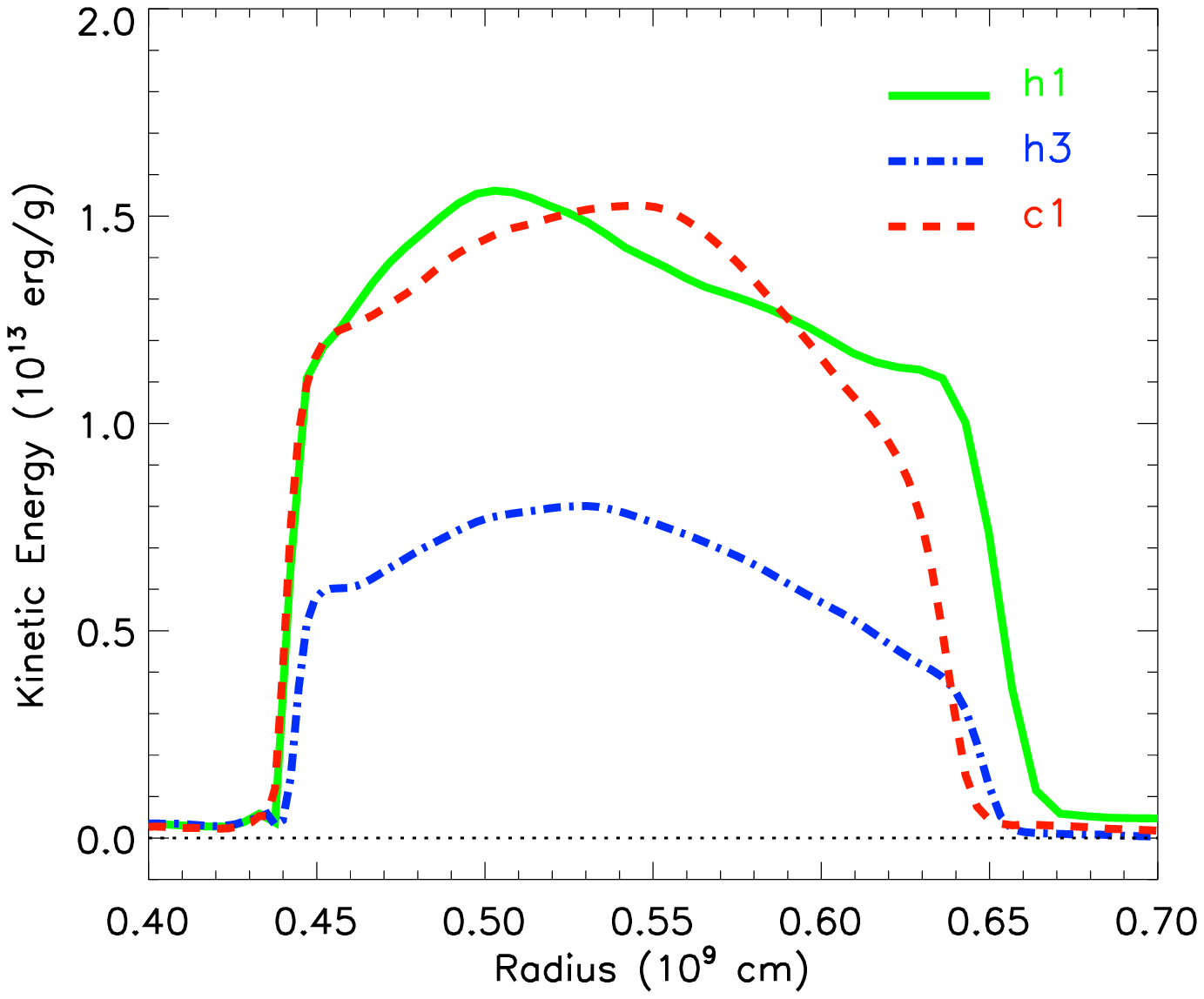}
\includegraphics[scale=0.52]{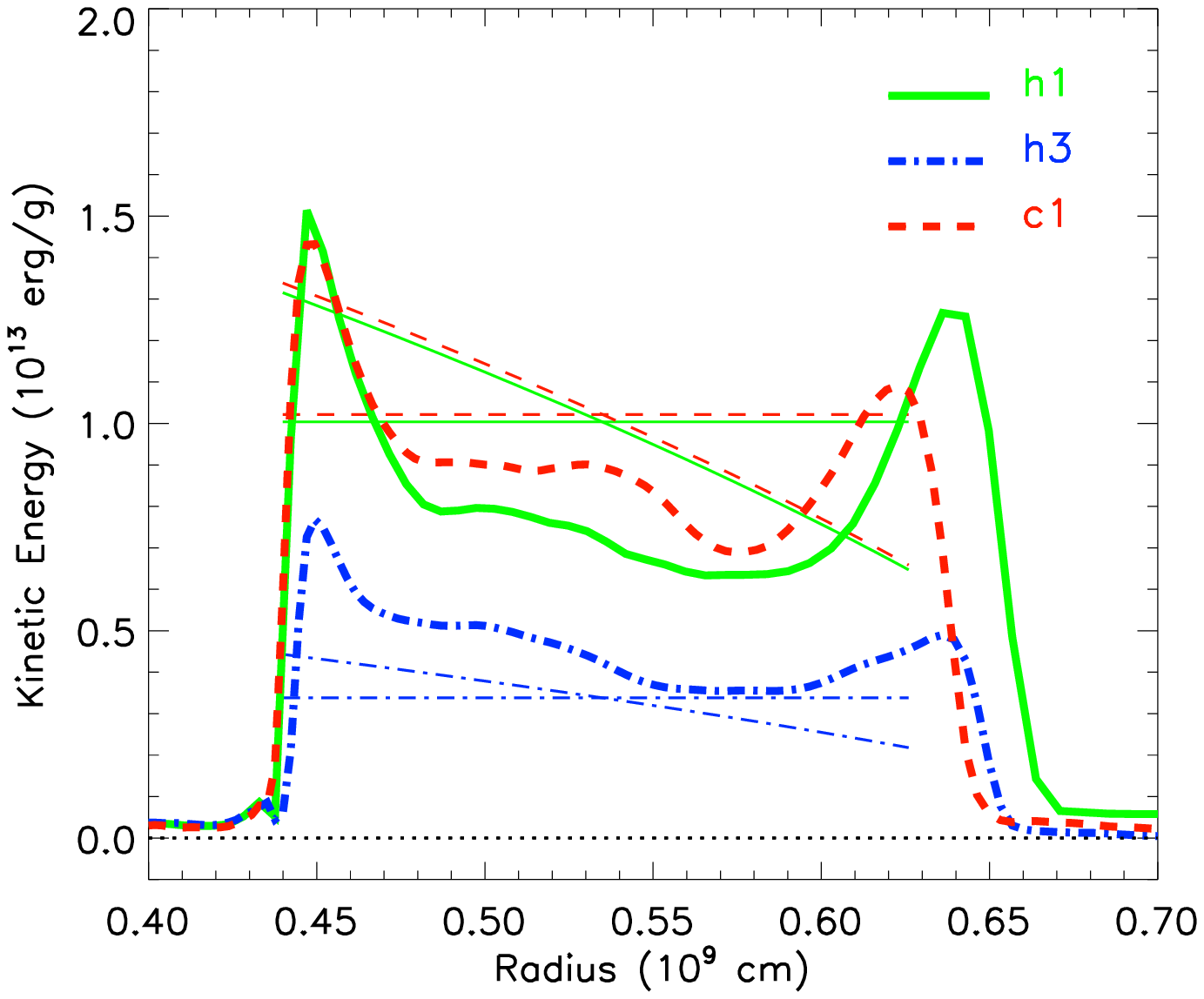}
\caption{Kinetic energy profiles: (left) total value of $E_K$ and (right) effective isotropic value $E_{K,{\rm iso}}$ described in \S4. The line segments indicate the two assumed distributions also described in \S4.}
\end{figure*}

\begin{figure*}
\includegraphics[scale=0.52]{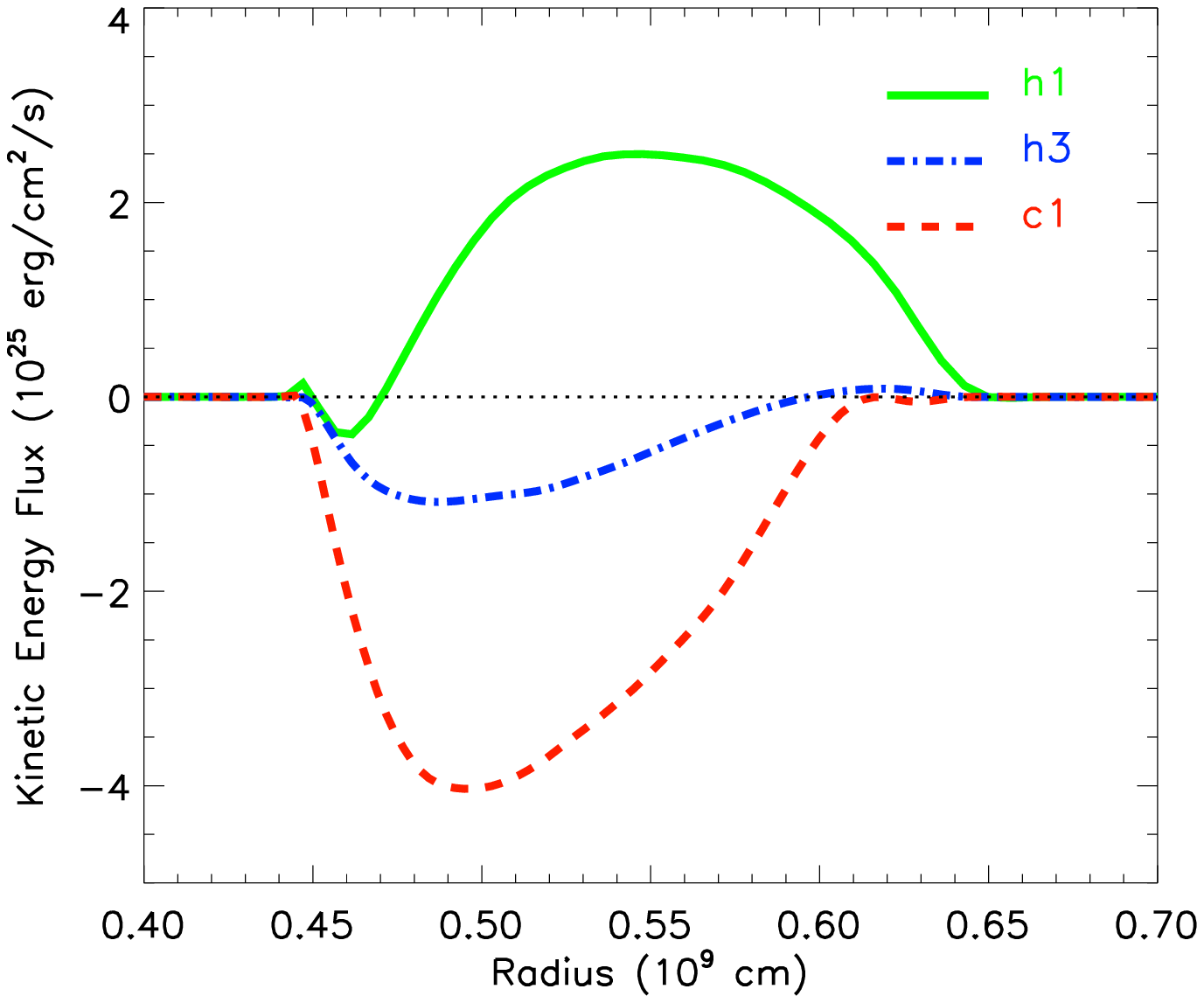}
\includegraphics[scale=0.52]{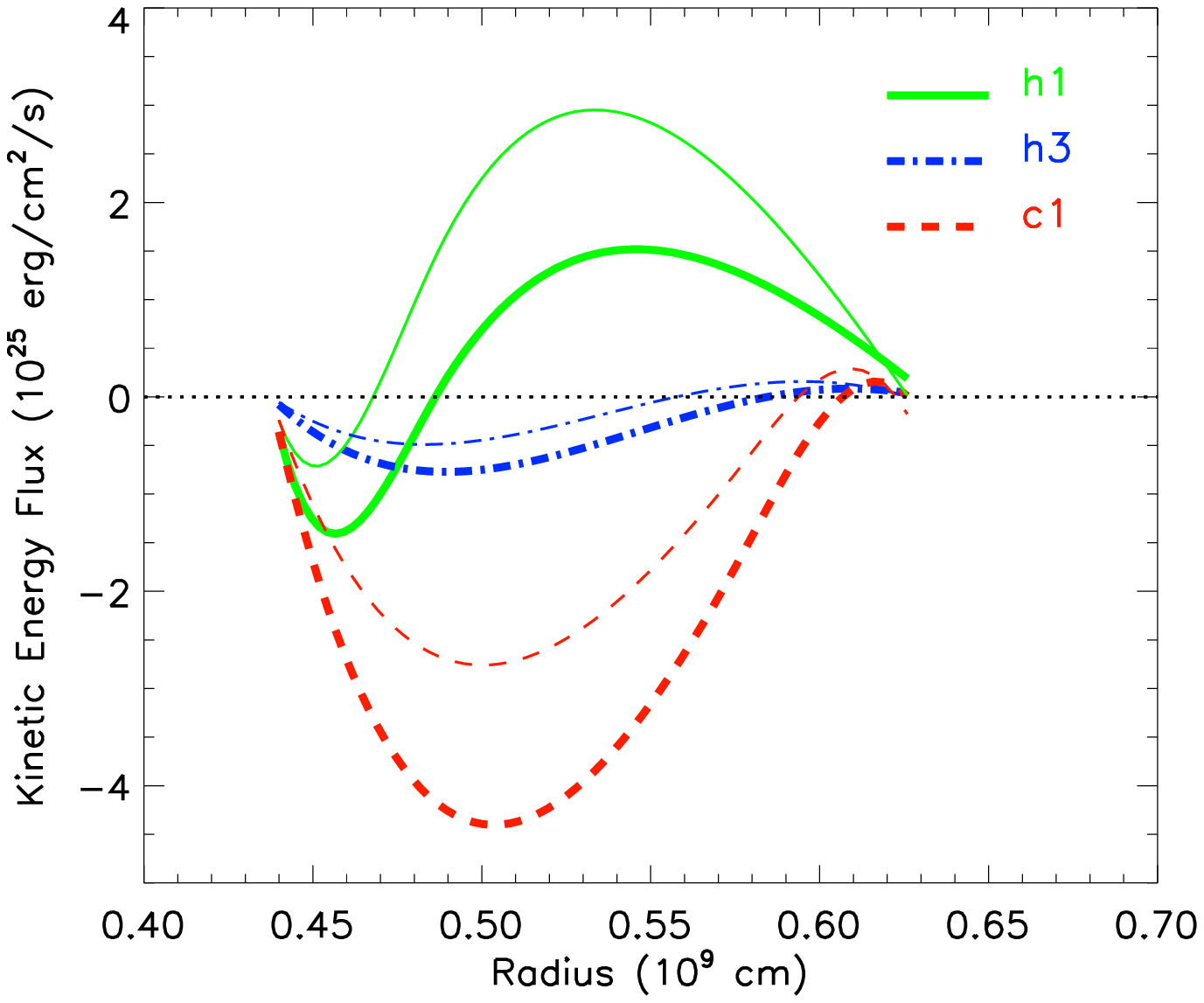}
\caption{Kinetic energy flux: (left) time averaged simulation data and
(right) values calculated directly from background structure using Eq. 4. The line thickness in the theoretical profiles indicate the assumed distribution of $E_{K,{\rm iso}}$ shown in Fig. 3: the (thin line) is the uniform case and the (thick line) the case with a gradient. }
\end{figure*}

\section{Kinetic Energy Flux and $\epsilon_K$}
\label{sec:keflux}

\par In quasi-steady states where $\vec{F}_p$ and $p'\nabla\cdot{\vec{u'}}$ are not important the kinetic energy flux (or luminosity) can be found by integrating Eq.~1

\begin{equation}\label{eq:ltot}
L_K(m) = \int_{M_0}^{M_0+m} \Big(L_C\nabla_{ad}\frac{dr'}{H_p} -\epsilon_K dm' \Big)
\end{equation}

\noindent with $dr' = dm'/4\pi r'^2\rho$. In this expression the radial component of the rate of buoyancy work is written in terms of the convective energy flux with $W_b = F_C\nabla_{ad}/H_p$ for pressure scale height $H_p$ and adiabatic gradient $\nabla_{ad} = (d\ln T/d\ln P)_s$.  This expression for $W_b$ can be calculated directly from the background structure (Eq.~3) and is appropriate for small density fluctuations that can be linearly related to temperature fluctuations using the isobaric thermodynamic derivative, a good approximation in most cases of deep, nearly adiabatic convection.  From Eq.~\ref{eq:ltot} we see that {\em the kinetic energy flux is the residual between buoyancy driving and viscous dissipation.}   

\par Globally, the integrated dissipation $\int\epsilon_K dm$ is constrained by both the thermal state evolution, $T\dot{s}$ (Eq. 2, 3), and the balance with buoyancy driving (Eq. 4, noting that $L_K(r_{\rm top}) = L_K(r_{\rm bot}) = 0$).  The radial profile of $\epsilon_K$ is determined by the topology of the convective flow.  \citet{arnett2009a} show that the dissipation is well described by the properties of the {\rm isotropic} component of turbulence, $v_{\rm iso}^2 \sim \frac{3}{2}(v_{\theta}^2 + v_{\phi}^2)$ with $\epsilon_K \sim v_{\rm iso}^3/l_d$ where $l_d$ is the largest scale of motion in the flow and $v_{\theta}$ and $v_{\phi}$ are the non-radial velocity fluctuations. 

\par In Fig.~3 we present the radial distribution of the kinetic energy from the simulation data.
The first panel shows the total $E_K$ and the second panel shows the horizontal component scaled to an equivalent isotropic value, $E_{K,{\rm iso}} = \frac{3}{2}E_{K,H}$. The increase in $E_{K,H}$ at the boundaries of the convection zones are due to the horizontal deflection of the large scale flow and wave motions excited in stable layers \citep[e.g.][]{meakin2006,meakin2007a} and should be corrected for when identifying $E_{K,{\rm iso}}$ with the convective turbulence.  

\par In Fig.~3 (right) we over plot two approximations to $E_{K,{\rm iso}}$: one based on a uniform distribution of dissipation and one based on a dissipation that decreases linearly with enclosed mass. The relationship $\epsilon_K = (2 E_{K,{\rm iso}})^{3/2}/l_d$ with $l_d = H_p$ is used.  The absolute scale of the dissipation and kinetic energy profiles are provided by the constraint that the global dissipation rate must balance the global rate of buoyancy driving.  The amplitude of the kinetic energy that satisfies this global balance is found by varying it until the boundary conditions on $L_K$ are satisfied (i.e., $L_K=0$ at the boundaries of the convection zone).  $E_{K,{\rm iso}} \sim \frac{1}{2}(F_c/\rho)^{2/3}$ provides a good first approximation. 

\par The kinetic energy fluxes found using this procedure are compared to the simulation data in Fig.~4 for the two assumed dissipation profiles.  

\section{Discussion}

\par We have provided a basic overview of the connection between turbulent dissipation and the kinetic energy flux in efficient (high P\'{e}clet number) convection. The only assumption made in our analysis involved the radial profile of the dissipation $\epsilon_K$ which we will discuss in a future publication. For now we shall suffice to say that the dissipation can be derived directly from the stellar model by adopting certain constraints on the topology of the convective flow. In particular, a two component flow model consisting of a background isotropic turbulent state and a large scale, plume-like flow is a promising approach.

\par  The data presented in Figs.~3 and 4 illustrate the shortcomings of the commonly used closure relation referred to as the {\em down gradient approximation}\footnote{The down gradient approximation is a closure relationship which relates the kinetic energy flux to the gradient of the turbulent kinetic energy such that $L_K \propto -\nabla E_K$.}. 
Most illustrative is the fact that while the $E_K$ distributions are nearly identical in models {\sf h1} and {\sf c1} the $L_K$ profiles are roughly mirror images.  The locally defined down gradient approximation flux fails because the properties of the turbulent transport are strongly shaped by global constraints, a feature that is captured by the analysis presented in \S\ref{sec:keflux}.

\par Another consequence of the distribution of kinetic energy within the convection zone is the rate of boundary layer mixing (see Fig. 2), which can significantly modify the stellar structure on evolutionary timescales \citep[see \S7 in][]{meakin2007b}. 

\acknowledgments

This work was supported by NSF Grant 0708871 and NASA Grant NNX08AH19G at the University of Arizona.
We thank Frank Timmes for generously providing computing hours on the {Saguaro} system at Arizona State University and Douglas Fuller for computer support.

\bibliographystyle{spr-mp-nameyear-cnd}

\end{document}